\documentclass[onecolumn]{aa}

\usepackage{amsmath,amssymb}
\usepackage{graphicx}
\usepackage{txfonts}
\usepackage{natbib}

\begin{document}

\title{Isotropic inelastic and superelastic collisional rates \\
in a multiterm atom}

\titlerunning{Inelastic and superelastic collisional rates in a multiterm atom}

\author{
L. Belluzzi\inst{\ref{inst1},\ref{inst2}}
\and E. Landi Degl'Innocenti\inst{\ref{inst3}} 
\and J. Trujillo Bueno\inst{\ref{inst1},\ref{inst2},\ref{inst4}}
}

\authorrunning{Belluzzi, Landi Degl'Innocenti, \& Trujillo Bueno}

\institute{
Instituto de Astrof\'isica de Canarias, C. V\'ia L\'actea s/n, E-38205 La 
Laguna, Tenerife, Spain\label{inst1}
\and
Departamento de Astrof\'isica, Facultad de F\'isica, Universidad de La Laguna, 
E-38200 La Laguna, Tenerife, Spain\label{inst2}
\and
Dipartimento di Fisica e Astronomia, Sezione di Astronomia e Scienza dello 
Spazio, Universit\`a di Firenze, Largo E. Fermi 2, I-50125 Firenze, 
Italy\label{inst3}
\and
Consejo Superior de Investigaciones Cient\'ificas, Spain\label{inst4}
}

\abstract{
The spectral line polarization of the radiation emerging from a magnetized
astrophysical plasma depends on the state of the atoms within the medium,
whose determination requires considering the interactions between the atoms 
and the magnetic field, between the atoms and photons (radiative transitions), 
and between the atoms and other material particles (collisional transitions).
In applications within the framework of the multiterm model atom (which 
accounts for quantum interference between magnetic sublevels pertaining either 
to the same $J$-level or to different $J$-levels within the same term) 
collisional processes are generally neglected when solving the master equation 
for the atomic density matrix.
This is partly due to the lack of experimental data and/or of approximate 
theoretical expressions for calculating the collisional transfer and 
relaxation rates (in particular the rates for interference between sublevels 
pertaining to different $J$-levels, and the depolarizing rates due to elastic 
collisions).
In this paper we formally define and investigate the transfer and relaxation 
rates due to isotropic inelastic and superelastic collisions that enter the 
statistical equilibrium equations for the atomic density matrix of a 
multiterm atom.
Under the hypothesis that the interaction between the collider and the atom can 
be described by a dipolar operator, we provide expressions that relate the 
collisional rates for interference between different $J$-levels to the usual 
collisional rates for $J$-level populations, for which experimental data or 
approximate theoretical expressions are generally available.
We show that the rates for populations and interference within the same
$J$-level reduce to those previously obtained for the multilevel model atom
(where quantum interference is assumed to be present only between magnetic
sublevels pertaining to any given $J$-level).
Finally, we apply the general equations to the case of a two-term atom with 
unpolarized lower term, illustrating the impact of inelastic and
superelastic collisions on the scattering line polarization through radiative 
transfer calculations in a slab of stellar atmospheric plasma anisotropically 
illuminated by the photospheric radiation field.
}


\maketitle

\section{Introduction}
The intensity and polarization of the spectral line radiation emerging from an 
astrophysical plasma depends on the population and atomic polarization (i.e., 
population imbalances and quantum interference between different magnetic 
sublevels) of the lower and upper line levels at each spatial point along the 
line of sight (LOS).
Determining the population and atomic polarization of such levels requires 
considering the interactions between the atoms and photons (radiative 
transitions) and between the atoms and other material particles, such as 
electrons, atoms, and ions (collisional transitions).
This problem can be very complex, especially when it comes to modeling the 
spectral line polarization produced by the joint action of anisotropic 
radiation pumping and the Hanle and Zeeman effects in multilevel systems. 

Within the framework of the density-matrix theory of spectral line polarization 
described in the monograph by Landi Degl'Innocenti \& Landolfi (2004; hereafter 
LL04), it is possible to develop a consistent set of equations for multilevel 
systems, either by neglecting (multilevel model atom) or considering
(multiterm model atom) quantum interference between pairs of magnetic 
sublevels pertaining to different $J$-levels (with $J$ the level's total 
angular momentum value). 
The relevant equations are the radiative transfer equation for the Stokes 
parameters (where the coefficients of the emission vector and of the 
propagation matrix depend on the values of the atomic density matrix) and the 
master equation for the atomic density matrix (which includes both radiative 
and collisional rates). 

While for the multilevel model atom LL04 derived the expressions for both 
radiative and collisional rates (assuming isotropic collisions), for the 
multiterm model atom they only provide the expressions for the radiative 
rates. 
The aim of this paper is to formally define the collisional rates for a 
multiterm atom, and to find their relevant properties, focusing our attention 
only on isotropic inelastic and superelastic collisions. 
The treatment of elastic collisions in a multiterm atom is actually more
complicated, and will not be treated here. 
Such collisions (e.g., with neutral hydrogen atoms) tend to equalize the 
populations of the sublevels pertaining to any given $J$-level and to destroy 
any quantum interference between pairs of them. 
A similar depolarizing role may be caused by inelastic and superelastic 
collisions between the $J$-levels pertaining to any given term, especially 
when such $J$-levels are very close in energy \citep[see][for the hydrogen 
case]{Bom09}.
For the sake of simplicity, the latter type of collision will also be 
neglected, the investigation being limited to inelastic and superelastic 
collisions between different terms.

In the main body of this paper, we provide suitable expressions for the 
transfer and relaxation rates caused by isotropic inelastic and superelastic 
collisions, taking the possibility of atomic polarization in all the terms of 
the model atom into account.
Particular attention is given to the collisional transfer and relaxation rates 
for interference between magnetic sublevels pertaining to different $J$-levels, 
the physical ingredient that cannot be accounted for with a multilevel model 
atom. 
Since there are basically no experimental data for such rates, we provide 
approximate expressions here that relate such rates to the usual collisional 
rates that describe transitions between different $J$-levels (for which 
experimental data or theoretical expressions are generally available).
As a consistency proof of our derivation, we show that the transfer and 
relaxation rates for populations and interference between pairs of magnetic 
sublevels pertaining to the same $J$-level reduce to those derived by LL04 for 
the multilevel atom case. 

In the last section we present an illustrative application of the theoretical 
scheme developed here. We consider a two-term atom with unpolarized lower term, 
and we show the sensitivity to the collisional rates of the linear polarization 
of the radiation emerging from a slab of given optical depth, located at a 
given height above the ``surface" of a solar-like star, and illuminated by its 
photospheric radiation field.

\section{Transfer rate due to inelastic collisions}
\label{Sect:CI}
We consider a multiterm atom (see Sect. 7.5 and 7.6 of LL04) in the absence of 
magnetic fields, and we describe it by means of the density matrix elements 
$\rho_{\beta L S}(J M, J^{\prime} M^{\prime})$, with $J$ the total angular 
momentum, $M$ its projection along the quantization axis, $L$ the orbital 
angular momentum, $S$ the spin, and $\beta$ the electronic configuration.
This atomic model accounts for quantum interference (or coherence) between 
pairs of magnetic sublevels pertaining either to the same $J$-level or to 
different $J$-levels of the same term ($J$-state interference).
We also work using the multipole moments of the density matrix (or spherical 
statistical tensors), defined by the equation
\begin{equation}
	^{\beta L S} \! \rho^K_Q(J,J^{\prime}) = \sum_{M M^{\prime}} 
	(-1)^{J-M} \sqrt{2K+1}
	\left( \begin{array}{ccc}
		J & J^{\prime} & K \\
		M & -M^{\prime} & -Q 
	\end{array} \right)
	\rho_{\beta L S}(JM, J^{\prime} M^{\prime}) \; .
\label{Eq:rho_KQ}
\end{equation}
Although collisional processes can be very efficient in coupling $J$-levels 
pertaining to the same term, in this investigation we only consider collisional 
processes coupling populations and coherence pertaining to different terms.

In a given, although arbitrary, reference system, transfer processes due to 
inelastic collisions contribute to the time evolution of a particular density 
matrix element according to the equation
\begin{equation}
	\frac{\rm d}{{\rm d} t} \, 
	\rho_{\beta L S}(J M, J^{\prime}M^{\prime}) =
	\sum_{\beta_{\ell} L_{\ell} J_{\ell} M_{\ell} J_{\ell}^{\prime} 
		M_{\ell}^{\prime}}
	C_I(\beta L S J M J^{\prime} M^{\prime}, \beta_{\ell} L_{\ell} S 
		J_{\ell} M_{\ell} J_{\ell}^{\prime} M_{\ell}^{\prime}) \, 
	\rho_{\beta_{\ell} L_{\ell} S}(J_{\ell} M_{\ell}, J_{\ell}^{\prime} 
		M_{\ell}^{\prime}) \; ,
\label{Eq:CI_std}
\end{equation}
where $C_I$ is the inelastic collision transfer rate and where the quantum 
numbers $(\beta_{\ell} L_{\ell} S)$ denote any term having energy lower than 
the term $(\beta L S)$.\footnote{We assume that there is no overlapping in 
energy among the various terms of the model atom under consideration.}
In a new reference system, obtained from the old one by the rotation $R$, 
recalling the transformation law (see Eq.~(3.95) of LL04)
\begin{equation}
	\left[ \rho_{\beta L S}(J M, J^{\prime} M^{\prime}) \right]_{\rm new} =
	\sum_{N N^{\prime}} {\mathcal D}^J_{NM}(R)^{\ast} \,
	{\mathcal D}^{J^{\prime}}_{N^{\prime} M^{\prime}}(R) \left[ 
	\rho_{\beta L S}(J N, J^{\prime} N^{\prime}) \right]_{\rm old} \; ,
\label{Eq:rot1}
\end{equation}
with ${\mathcal D}^{J}_{M N}(R)$ the rotation matrices, and its inverse
\begin{equation}
	\left[ \rho_{\beta L S}(J M, J^{\prime} M^{\prime}) \right]_{\rm old} =
	\sum_{N N^{\prime}} {\mathcal D}^J_{M N}(R) \,
	{\mathcal D}^{J^{\prime}}_{M^{\prime} N^{\prime}}(R)^{\ast} \left[ 
	\rho_{\beta L S}(J N, J^{\prime} N^{\prime}) \right]_{\rm new} \; ,
\label{Eq:rot2}
\end{equation}
we have
\begin{equation}
\begin{split}
	\frac{\rm d}{{\rm d} t} 
	\left[ \rho_{\beta L S}(J M, J^{\prime} M^{\prime}) \right]_{\rm new} =
	\sum_{\beta_{\ell} L_{\ell} J_{\ell} M_{\ell} J_{\ell}^{\prime} 
		M_{\ell}^{\prime}} &
	\left\{ \sum_{N N_{\phantom{\ell}}^{\prime} N_{\ell} N_{\ell}^{\prime}} 
	{\mathcal D}^{J}_{N M}(R)^{\ast} \,
	{\mathcal D}^{J^{\prime}}_{N^{\prime} M^{\prime}}(R) \,
	{\mathcal D}^{J_{\ell}}_{N_{\ell} M_{\ell}}(R) \,
	{\mathcal D}^{J^{\prime}_{\ell}}_{N^{\prime}_{\ell} 
		M^{\prime}_{\ell}}(R)^{\ast} \,
	C_I(\beta L S J N J^{\prime} N^{\prime}, \beta_{\ell} L_{\ell} S 
	    J_{\ell} N_{\ell} J_{\ell}^{\prime} N_{\ell}^{\prime}) \right\} \\
	& \times \, \left[ \rho_{\beta_{\ell} L_{\ell} S}(J_{\ell} M_{\ell}, 
		J_{\ell}^{\prime} M_{\ell}^{\prime}) \right]_{\rm new} \; .
\label{Eq:CI_std_new}
\end{split}
\end{equation}
The assumption of isotropic collisions implies that all the quantization 
directions are equivalent, so that Eqs.~(\ref{Eq:CI_std}) and 
(\ref{Eq:CI_std_new}) have to be identical.
It follows that the collisional rates must satisfy the relation
\begin{equation}
	C_I(\beta L S J M J^{\prime} M^{\prime}, \beta_{\ell} L_{\ell} S 
		J_{\ell} M_{\ell} J_{\ell}^{\prime} M_{\ell}^{\prime}) = 
	\sum_{N N_{\phantom{\ell}}^{\prime} N_{\ell} N_{\ell}^{\prime}} 
	{\mathcal D}^{J}_{N M}(R)^{\ast} \,
	{\mathcal D}^{J^{\prime}}_{N^{\prime} M^{\prime}}(R) \,
	{\mathcal D}^{J_{\ell}}_{N_{\ell} M_{\ell}}(R) \,
	{\mathcal D}^{J^{\prime}_{\ell}}_{N^{\prime}_{\ell} 
		M^{\prime}_{\ell}}(R)^{\ast} \,
	C_I(\beta L S J N J^{\prime} N^{\prime}, \beta_{\ell} L_{\ell} S 
		J_{\ell} N_{\ell} J_{\ell}^{\prime} N_{\ell}^{\prime}) \; .
\label{Eq:CI_rel1}
\end{equation}
After coupling through Eq.~(\ref{Eq:rotmat1}) the rotation matrices 
${\mathcal D}^{J}_{N M}(R)^{\ast}$ and 
${\mathcal D}^{J^{\prime}}_{N^{\prime} M^{\prime}}(R)$, as well as the rotation 
matrices ${\mathcal D}^{J_{\ell}}_{N_{\ell} M_{\ell}}(R)^{\ast}$ and 
${\mathcal D}^{J^{\prime}_{\ell}}_{N^{\prime}_{\ell} M^{\prime}_{\ell}}(R)$, 
and using the complex conjugate of Eq.~(\ref{Eq:rotmat2}) on the ensuing 
expression, Eq.~(\ref{Eq:CI_rel1}) takes the form
\begin{equation}
\begin{split}
	C_I(\beta L S J M J^{\prime} M^{\prime}, \beta_{\ell} L_{\ell} S 
		J_{\ell} M_{\ell} J_{\ell}^{\prime} M_{\ell}^{\prime}) = & 
	\sum_{N N_{\phantom{\ell}}^{\prime} N_{\ell} N_{\ell}^{\prime}}
	C_I(\beta L S J N J^{\prime} N^{\prime}, \beta_{\ell} L_{\ell} S 
		J_{\ell} N_{\ell} J_{\ell}^{\prime} N_{\ell}^{\prime}) \\
	& \times \, \sum_{K K^{\prime} K^{\prime \prime}} 
	(-1)^{N - M + N_{\ell}^{\prime} - M_{\ell}^{\prime}}
	(2K + 1) (2K^{\prime} + 1) (2K^{\prime \prime} + 1) \\
	& \times \, 
	\left( \begin{array}{ccc}
		J^{\prime} & J & K \\
		N^{\prime} & -N & P 
	\end{array} \right)
	\left( \begin{array}{ccc}
		J^{\prime} & J & K \\
		M^{\prime} & -M & Q 
	\end{array} \right)
	\left( \begin{array}{ccc}
		J_{\ell} & J_{\ell}^{\prime} & K^{\prime} \\
		N_{\ell} & -N_{\ell}^{\prime} & P^{\prime} 
	\end{array} \right)
	\left( \begin{array}{ccc}
		J_{\ell} & J_{\ell}^{\prime} & K^{\prime} \\
		M_{\ell} & -M_{\ell}^{\prime} & Q^{\prime} 
	\end{array} \right) \\
	& \times \, 
	\left( \begin{array}{ccc}
		K & K^{\prime} & K^{\prime \prime} \\
		P & P^{\prime} & P^{\prime \prime} 
	\end{array} \right)
	\left( \begin{array}{ccc}
		K & K^{\prime} & K^{\prime \prime} \\
		Q & Q^{\prime} & Q^{\prime \prime} 
	\end{array} \right)
	{\mathcal D}^{K^{\prime \prime}}_{P^{\prime \prime} 
	Q^{\prime \prime}}(R) \; .
\label{Eq:CI_rel2}
\end{split}
\end{equation}
As the righthand side of Eq.~(\ref{Eq:CI_rel2}) must be independent of 
the rotation $R$, the index $K^{\prime \prime}$ can only take the value 
$K^{\prime \prime}=0$. This implies $K=K^{\prime}$, $P=-P^{\prime}$, and
$Q=-Q^{\prime}$. Using Eq.~(\ref{Eq:3jc}), we obtain
\begin{equation}
\begin{split}
	C_I(\beta L S J M J^{\prime} M^{\prime}, \beta_{\ell} L_{\ell} S 
		J_{\ell} M_{\ell} J_{\ell}^{\prime} M_{\ell}^{\prime}) = & 
	\sum_{N N_{\phantom{\ell}}^{\prime} N_{\ell} N_{\ell}^{\prime}}
	C_I(\beta L S J N J^{\prime} N^{\prime}, \beta_{\ell} L_{\ell} S 
		J_{\ell} N_{\ell} J_{\ell}^{\prime} N_{\ell}^{\prime}) \,
	\sum_K (2K + 1) \, 
	(-1)^{N - M + N_{\ell}^{\prime} - M_{\ell}^{\prime} - P - Q} \\
	& \times \,
	\left( \begin{array}{ccc}
		J^{\prime} & J & K \\
		N^{\prime} & -N & P 
	\end{array} \right)
	\left( \begin{array}{ccc}
		J^{\prime} & J & K \\
		M^{\prime} & -M & Q 
	\end{array} \right)
	\left( \begin{array}{ccc}
		J_{\ell} & J_{\ell}^{\prime} & K \\
		N_{\ell} & -N_{\ell}^{\prime} & -P
	\end{array} \right)
	\left( \begin{array}{ccc}
		J_{\ell} & J_{\ell}^{\prime} & K \\
		M_{\ell} & -M_{\ell}^{\prime} & -Q 
	\end{array} \right)
	\; .
\label{Eq:CI_rel3}
\end{split}
\end{equation}

\subsection{Multipole components of the inelastic collision transfer rate}
Introducing the multipole components of the inelastic collision transfer rate, 
defined by the equation\footnote{The factor 
$\sqrt{J + J^{\prime}+1}/\sqrt{J_{\ell} + J_{\ell}^{\prime}+1}$ is 
introduced in order to get simpler relations between these rates and the usual 
collisional rates connecting atomic populations. This factor reduces to the one 
introduced in the multilevel atom case (see Eq.~(7.87) of LL04) when 
interference between different $J$-levels is neglected ($J=J^{\prime}$ and
$J_{\ell}=J_{\ell}^{\prime}$).}
\begin{equation}
\begin{split}
	C_I^{(K)}(\beta L S J J^{\prime}, \beta_{\ell} L_{\ell} S J_{\ell} 
		J_{\ell}^{\prime}) = &
	\sqrt{\frac{J + J^{\prime} + 1}{J_{\ell} + J_{\ell}^{\prime} + 1}} \\
	& \times \sum_{N N^{\prime} N_{\ell} N_{\ell}^{\prime}} 
	(-1)^{J + J_{\ell} - N^{\prime} - N_{\ell}^{\prime}}
	\left( \begin{array}{ccc}
		J^{\prime} & J & K \\
		N^{\prime} & -N & P 
	\end{array} \right)
	\left( \begin{array}{ccc}
		J_{\ell}^{\prime} & J_{\ell} & K \\
		N_{\ell}^{\prime} & -N_{\ell} & P
	\end{array} \right)
	C_I(\beta L S J N J^{\prime} N^{\prime}, \beta_{\ell} L_{\ell} S 
		J_{\ell} N_{\ell} J_{\ell}^{\prime} N_{\ell}^{\prime}) \, ,
\label{Eq:CIK_def}
\end{split}
\end{equation}
and making use of Eqs.~(\ref{Eq:3ja}) and (\ref{Eq:3jb}), 
Eq.~(\ref{Eq:CI_rel3}) can be written in the form
\begin{equation}
\begin{split}
	C_I(\beta L S J M J^{\prime} M^{\prime}, \beta_{\ell} L_{\ell} S 
		J_{\ell} M_{\ell} J_{\ell}^{\prime} M_{\ell}^{\prime}) = & 
	\sqrt{\frac{J_{\ell} + J_{\ell}^{\prime} + 1}{J + J^{\prime} + 1}} \,
	(-1)^{J + J_{\ell} - M^{\prime} - M_{\ell}^{\prime}} \\
	& \times \sum_K (2K + 1) 
	\left( \begin{array}{ccc}
		J^{\prime} & J & K \\
		M^{\prime} & -M & Q 
	\end{array} \right)
	\left( \begin{array}{ccc}
		J_{\ell}^{\prime} & J_{\ell} & K \\
		M_{\ell}^{\prime} & -M_{\ell} & Q 
	\end{array} \right)
	C_I^{(K)}(\beta L S J J^{\prime}, \beta_{\ell} L_{\ell} S J_{\ell} 
		J_{\ell}^{\prime}) \; .
\label{Eq:CI_CIK_rel}
\end{split}
\end{equation}
Substituting Eq.~(\ref{Eq:CI_CIK_rel}) into Eq.~(\ref{Eq:CI_std}), and 
recalling the definition of the multipole moments of the density matrix 
(see Eq.~(\ref{Eq:rho_KQ})), with the help of Eq.~(\ref{Eq:3j_orto}), we find 
the following equation for the spherical statistical tensors
\begin{equation}
	\frac{\rm d}{{\rm d} t} \, 
	^{\beta L S} \! \rho^K_Q(J, J^{\prime}) = 
	\sum_{\beta_{\ell} L_{\ell} J_{\ell} J_{\ell}^{\prime}}
	\sqrt{\frac{J_{\ell} + J_{\ell}^{\prime} + 1}{J + J^{\prime} + 1}} \,
	C_I^{(K)}(\beta L S J J^{\prime}, \beta_{\ell} L_{\ell} S J_{\ell} 
		J_{\ell}^{\prime}) \;
	^{\beta_{\ell} L_{\ell} S} \! \rho^K_Q(J_{\ell}, J_{\ell}^{\prime}) \; .
\end{equation}
Taking the complex conjugate of Eq.~(\ref{Eq:CI_std}) and recalling that 
$\rho_{\beta L S}(J M, J^{\prime} M^{\prime})^{\ast} = \rho_{\beta L S}
(J^{\prime} M^{\prime}, J M)$, we have
\begin{equation}
	C_I(\beta L S J M J^{\prime} M^{\prime}, \beta_{\ell} L_{\ell} S 
		J_{\ell} M_{\ell} J_{\ell}^{\prime} M_{\ell}^{\prime})^{\ast}=  
	C_I(\beta L S J^{\prime} M^{\prime} J M, \beta_{\ell} L_{\ell} S 
	J_{\ell}^{\prime} M_{\ell}^{\prime} J_{\ell} M_{\ell}) \; ,
\end{equation}
and therefore, using Eqs.~(\ref{Eq:3ja}) and (\ref{Eq:3jb}),
\begin{equation}
	C_I^{(K)}(\beta L S J J^{\prime}, \beta_{\ell} L_{\ell} S J_{\ell} 
		J_{\ell}^{\prime})^{\ast} =
	(-1)^{J + J_{\ell} - J^{\prime} - J_{\ell}^{\prime}} \, 
	C_I^{(K)}(\beta L S J^{\prime} J, \beta_{\ell} L_{\ell} S 
		J_{\ell}^{\prime} J_{\ell}) \; .
\end{equation}
Setting $K=0$ in Eq.~(\ref{Eq:CIK_def}), and using Eq.~(\ref{Eq:3jc}),
we obtain
\begin{equation}
	C_I^{(0)}(\beta L S J J^{\prime}, \beta_{\ell} L_{\ell} S J_{\ell} 
		J_{\ell}^{\prime}) =
	\delta_{J J^{\prime}} \, \delta_{J_{\ell} J_{\ell}^{\prime}} \,
	\frac{1}{2J_{\ell} +1} \sum_{N N_{\ell}}
	C_I(\beta L S J N J N, \beta_{\ell} L_{\ell} S J_{\ell} N_{\ell} 
		J_{\ell} N_{\ell}) \; ,
\label{Eq:CI0_1}
\end{equation}
where the transfer rate $C_I(\beta L S J N J N, \beta_{\ell} L_{\ell} S 
J_{\ell} N_{\ell} J_{\ell} N_{\ell})$ is the usual (inelastic) collisional 
rate for the transition from the lower magnetic sublevel 
$|\, \beta_{\ell} L_{\ell} S J_{\ell} N_{\ell} \rangle$ to the upper magnetic 
sublevel $|\, \beta L S J N \rangle$, generally indicated in the literature 
with the notation $\mathcal{C}_I(\beta_{\ell} L_{\ell} S J_{\ell} N_{\ell} 
\rightarrow \beta L S J N)$.
Since this rate is non-negative, the 0-rank multipole component is also
non-negative.

\subsection{Relations with the collisional rates for $J$-level populations}
In most cases, the only collisional rates for which experimental data, or 
approximate analytical expressions, are available are the collisional rates 
connecting the populations of different $J$-levels (following the notation 
generally used in the literature, these rates will be indicated through the 
symbols $\mathcal{C}_I(\beta_{\ell} L_{\ell} S J_{\ell} \rightarrow \beta_u 
L_u S J_u)$ and $\mathcal{C}_S(\beta_u L_u S J_u \rightarrow \beta_{\ell} 
L_{\ell} S J_{\ell})$, the indices $I$ and $S$ standing for ``inelastic'' and 
``superelastic'', respectively).
It is important therefore to find suitable relations between such rates and 
the collisional rates introduced in this paper for a multiterm atom.

Observing that
\begin{equation}
	\mathcal{C}_I(\beta_{\ell} L_{\ell} S J_{\ell} \rightarrow \beta_u L_u 
	S J_u) = \frac{1}{2J_{\ell} +1} \sum_{N_u N_{\ell}}
	\mathcal{C}_I(\beta_{\ell} L_{\ell} S J_{\ell} N_{\ell} \rightarrow 
	\beta_u L_u S J_u N_u) \; ,
\end{equation}
from Eq.~(\ref{Eq:CI0_1}) we immediately have
\begin{equation}
	C_I^{(0)}(\beta L S J J, \beta_{\ell} L_{\ell} S J_{\ell} J_{\ell}) =
	\mathcal{C}_I(\beta_{\ell} L_{\ell} S J_{\ell} \rightarrow \beta L S J) 
	\; .
\label{Eq:CI0_2}
\end{equation}
In Eq.~(\ref{Eq:CI_rel1}), if we couple through Eq.~(\ref{Eq:rotmat1}) the 
rotation matrices ${\mathcal D}^{J}_{N M}(R)^{\ast}$ and 
${\mathcal D}^{J_{\ell}}_{N_{\ell} M_{\ell}}(R)$, as well as the rotation 
matrices ${\mathcal D}^{J^{\prime}}_{N^{\prime} M^{\prime}}(R)$ and 
${\mathcal D}^{J^{\prime}_{\ell}}_{N^{\prime}_{\ell} M^{\prime}_{\ell}}
(R)^{\ast}$, by requiring that the ensuing expression is independent of the 
rotation $R$, we find the relation
\begin{equation}
\begin{split}
	C_I(\beta L S J M J^{\prime} M^{\prime}, \beta_{\ell} L_{\ell} S 
		J_{\ell} M_{\ell} J_{\ell}^{\prime} M_{\ell}^{\prime}) = & 
	\sum_{N N_{\phantom{\ell}}^{\prime} N_{\ell} N_{\ell}^{\prime}}
	C_I(\beta L S J N J^{\prime} N^{\prime}, \beta_{\ell} L_{\ell} S 
		J_{\ell} N_{\ell} J_{\ell}^{\prime} N_{\ell}^{\prime}) \,
	\sum_K (2K + 1) \, (-1)^{N - M + N_{\ell}^{\prime} - M_{\ell}^{\prime} 
		- P - Q} \\
	& \times \,
	\left( \begin{array}{ccc}
		J & J_{\ell} & K \\
		-M & M_{\ell} & Q 
	\end{array} \right)
	\left( \begin{array}{ccc}
		J^{\prime} & J^{\prime}_{\ell} & K \\
		-M^{\prime} & M^{\prime}_{\ell} & Q 
	\end{array} \right)
	\left( \begin{array}{ccc}
		J & J_{\ell} & K \\
		-N & N_{\ell} & P
	\end{array} \right)
	\left( \begin{array}{ccc}
		J^{\prime} & J_{\ell}^{\prime} & K \\
		-N^{\prime} & N_{\ell}^{\prime} & P 
	\end{array} \right)
	\; .
\label{Eq:CI_rel4}
\end{split}
\end{equation}
Defining a different set of multipole components of the inelastic collision
transfer rate through the equation
\begin{equation}
\begin{split}
	\Gamma_I^{(K)}(\beta L S J J^{\prime}, \beta_{\ell} L_{\ell} S J_{\ell} 
		J_{\ell}^{\prime}) =
	\frac{(2K + 1)}{J_{\ell} + J_{\ell}^{\prime} + 1}
	\sum_{N N^{\prime} N_{\ell} N_{\ell}^{\prime}} 
	(-1)^{N_{\ell}^{\prime} - N_{\ell}}
	\left( \begin{array}{ccc}
		J & J_{\ell} & K \\
		-N & N_{\ell} & P 
	\end{array} \right)
	\left( \begin{array}{ccc}
		J^{\prime} & J_{\ell}^{\prime} & K \\
		-N^{\prime} & N_{\ell}^{\prime} & P
	\end{array} \right)
	C_I(\beta L S J N J^{\prime} N^{\prime}, \beta_{\ell} L_{\ell} S 
		J_{\ell} N_{\ell} J_{\ell}^{\prime} N_{\ell}^{\prime}) \, ,
\label{Eq:GIK_def}
\end{split}
\end{equation}
Eq.~(\ref{Eq:CI_rel4}) can be written in the form
\begin{equation}
\begin{split}
	C_I(\beta L S J M J^{\prime} M^{\prime}, \beta_{\ell} L_{\ell} S 
		J_{\ell} M_{\ell} J_{\ell}^{\prime} M_{\ell}^{\prime}) = & 
	(-1)^{M_{\ell}^{\prime} - M_{\ell}} (J_{\ell} + J_{\ell}^{\prime} +1)
	\sum_K
	\left( \begin{array}{ccc}
		J & J_{\ell} & K \\
		-M & M_{\ell} & Q 
	\end{array} \right)
	\left( \begin{array}{ccc}
		J^{\prime} & J_{\ell}^{\prime} & K \\
		-M^{\prime} & M_{\ell}^{\prime} & Q 
	\end{array} \right)
	\Gamma_I^{(K)}(\beta L S J J^{\prime}, \beta_{\ell} L_{\ell} S J_{\ell} 
		J_{\ell}^{\prime}) \; .
\label{Eq:CI_GIK_rel}
\end{split}
\end{equation}
As pointed out in LL04 for the multilevel atom case, this decomposition of the 
collisional rate has an interesting physical interpretation, because it shows 
that the interaction between the atomic system and the collider can be 
described by a sum of tensor operators of rank $K$ acting on the state vectors 
of the atom.
Starting from Eq.~(\ref{Eq:CIK_def}) and using Eq.~(\ref{Eq:3j_contr}), after 
some algebra the following relation between the multipole components 
$C_I^{(K)}$ and $\Gamma_I^{(K)}$ can be found:
\begin{equation}
	C_I^{(K)}(\beta L S J J^{\prime}, \beta_{\ell} L_{\ell} S J_{\ell} 
		J_{\ell}^{\prime}) =
	\sqrt{(J + J^{\prime} + 1)(J_{\ell} + J_{\ell}^{\prime} + 1)}
	\sum_{K^{\prime}} (-1)^{J^{\prime} +J_{\ell}^{\prime} - K^{\prime} + K}
	\left\{ \begin{array}{ccc}
		J^{\prime} & J & K \\
		J_{\ell} & J_{\ell}^{\prime} & K^{\prime}
	\end{array} \right\}
	\Gamma_I^{(K^{\prime})}(\beta L S J J^{\prime}, \beta_{\ell} L_{\ell} S 
		J_{\ell} J_{\ell}^{\prime}) \; .
\end{equation}
For the $K=0$ multipole component, using Eq.~(\ref{Eq:6ja}), we have (cf. 
Appendix~A4 of LL04)
\begin{equation}
	C_I^{(0)}(\beta L S J J, \beta_{\ell} L_{\ell} S J_{\ell} J_{\ell}) =
	\sum_{K} \Gamma_I^{(K)}(\beta L S J J, \beta_{\ell} L_{\ell} S J_{\ell} 
		J_{\ell}) \; .
\end{equation}
When the interaction can be described through just one operator of rank 
$\tilde{K}$, then
\begin{equation}
	C_I^{(K)}(\beta L S J J^{\prime}, \beta_{\ell} L_{\ell} S J_{\ell} 
		J_{\ell}^{\prime}) =
	\sqrt{(J + J^{\prime} + 1)(J_{\ell} + J_{\ell}^{\prime} + 1)}
	\, (-1)^{J^{\prime} + J_{\ell}^{\prime} - \tilde{K} + K}
	\left\{ \begin{array}{ccc}
		J^{\prime} & J & K \\
		J_{\ell} & J_{\ell}^{\prime} & \tilde{K}
	\end{array} \right\}
	\Gamma_I^{(\tilde{K})}(\beta L S J J^{\prime}, \beta_{\ell} L_{\ell} S 
		J_{\ell} J_{\ell}^{\prime}) \; .
\end{equation}
The multipole component of rank $K$ of the diagonal rates ($J=J^{\prime}$ and 
$J_{\ell}=J_{\ell}^{\prime}$) is thus related to the multipole component of 
rank 0 by the equation (cf. Appendix~A4 of LL04)
\begin{equation}
	C_I^{(K)}(\beta L S J J, \beta_{\ell} L_{\ell} S J_{\ell} J_{\ell}) =
	(-1)^{K} \frac{
	\left\{ \begin{array}{ccc}
		J & J & K \\
		J_{\ell} & J_{\ell} & \tilde{K}
	\end{array} \right\}
	}{
	\left\{ \begin{array}{ccc}
		J & J & 0 \\
		J_{\ell} & J_{\ell} & \tilde{K}
	\end{array} \right\} }
	C_I^{(0)}(\beta L S J J, \beta_{\ell} L_{\ell} S J_{\ell} J_{\ell}) =
	(-1)^{K} \frac{
	\left\{ \begin{array}{ccc}
		J & J & K \\
		J_{\ell} & J_{\ell} & \tilde{K}
	\end{array} \right\}
	}{
	\left\{ \begin{array}{ccc}
		J & J & 0 \\
		J_{\ell} & J_{\ell} & \tilde{K}
	\end{array} \right\} }
	\mathcal{C}_I(\beta_{\ell} L_{\ell} S J_{\ell} \rightarrow \beta L S J) 
	\; .
\label{Eq:CIK_diag1}
\end{equation}

A similar relation for the nondiagonal rates (describing the transfer of 
$J$-state interference due to inelastic collisions) cannot be obtained through 
symmetry arguments alone.
Such a relation can, however, be derived if some simplifying hypotheses on the 
interaction between the atoms and perturbers are introduced.
It is well known that in the case of electrons with much higher energy
than the threshold energy (i.e. under the so-called Born approximation), the 
Hamiltonian describing the electron-atom interaction depends on the dynamical 
variables of the atom only through the dipole operator (a tensor of rank 
$\tilde{K}=1$).
A collisional process in an optically allowed transition can thus be treated, 
in a first approximation, as a radiative transition, and the collisional rate 
can be expressed through the oscillator strength of the same transition 
\citep[e.g.][]{Sea62,Reg62}.

For more insight on the nondiagonal rates $C_I^{(K)}(\beta L S J 
J^{\prime}, \beta_{\ell} L_{\ell} S J_{\ell} J_{\ell}^{\prime})$, we assume 
that the electron-atom interaction is described by a dipolar operator, and we 
proceed by analogy with the multiterm atom radiative transfer rate due to 
absorption processes ($\mathbb{T}_A$).
Setting $K_r = 0$ (i.e. assuming an isotropic radiation field) in Eq.~(7.45a) 
of LL04, we have
\begin{equation}
\begin{split}
	\mathbb{T}_A(\beta L S K Q J J^{\prime}, \beta_{\ell} L_{\ell} S K Q 
	J_{\ell} J_{\ell}^{\prime}) = & (2L_{\ell} + 1) 
	(-1)^{1 + J^{\prime} + J_{\ell}^{\prime} + K}
	\sqrt{(2J + 1)(2J^{\prime} + 1)(2J_{\ell} + 1)(2J_{\ell}^{\prime} +1)}
	\\
	& \times \, 
	\left\{ \begin{array}{ccc}
		J & J_{\ell} & 1 \\
		J_{\ell}^{\prime} & J^{\prime} & K
	\end{array} \right\}
	\left\{ \begin{array}{ccc}
		L & L_{\ell} & 1 \\
		J_{\ell} & J & S
	\end{array} \right\}
	\left\{ \begin{array}{ccc}
		L & L_{\ell} & 1 \\
		J_{\ell}^{\prime} & J^{\prime} & S
	\end{array} \right\}
	B(\beta_{\ell} L_{\ell} S \rightarrow \beta L S) J^0_0 \; ,
\label{Eq:Ta_0}
\end{split}
\end{equation}
where $J^0_0$ is the angle-averaged incident radiation field, and 
$B(\beta_{\ell} L_{\ell} S \rightarrow \beta L S)$ is the Einstein coefficient 
for absorption from the lower term $(\beta_{\ell} L_{\ell} S)$ to the upper 
term $(\beta L S)$. 
We recall that this quantity is connected to the Einstein coefficients for the 
individual transitions between fine structure $J$-levels of the multiplet by 
the relation (see Eq.~(7.57a) of LL04)
\begin{equation}
	B(\beta_{\ell} L_{\ell} S J_{\ell} \rightarrow \beta L S J) = 
	(2L_{\ell} +1) (2J + 1) 
	\left\{ \begin{array}{ccc}
		L & L_{\ell} & 1 \\
		J_{\ell} & J & S
	\end{array} \right\}^2
	B(\beta_{\ell} L_{\ell} S \rightarrow \beta L S) \; ,
\label{Eq:B_rel1}
\end{equation}
which implies (using Eq.~(\ref{Eq:6j_sum}))\footnote{The sum appearing on the 
righthand side of Eq.~(\ref{Eq:multi_term_Blu}) does not depend on the 
particular $J$-level of the lower term that is considered.}
\begin{equation}
	B(\beta_{\ell} L_{\ell} S \rightarrow \beta L S) = \sum_{J} 
	B(\beta_{\ell} L_{\ell} S J_{\ell} \rightarrow \beta L S J) \; .
\label{Eq:multi_term_Blu}
\end{equation}

By analogy with Eq.~(\ref{Eq:multi_term_Blu}), we define an inelastic 
collisional rate for the transition from the lower to the upper term 
$\mathcal{C}_I(\beta_{\ell} L_{\ell} S \rightarrow \beta L S)$ through the 
equation
\begin{equation}
	\mathcal{C}_I(\beta_{\ell} L_{\ell} S \rightarrow \beta L S) =
	\sum_{J} \mathcal{C}_I(\beta_{\ell} L_{\ell} S J_{\ell} \rightarrow 
	\beta L S J) \; ,
\label{Eq:CI_term1}
\end{equation}
where the sum is extended to all the $J$-levels of the upper term to which a 
given $J$-level of the lower term can be connected through an electric dipole 
transition.
By analogy with Eq.~(\ref{Eq:Ta_0}), and taking the multiplying factor 
introduced in Eq.~(\ref{Eq:CIK_def}) into account (see footnote 2), we can 
write
\begin{equation}
\begin{split}
	C_I^{(K)}(\beta L S J J^{\prime},\beta_{\ell} L_{\ell} S J_{\ell} 
		J_{\ell}^{\prime}) 
	= & (2L_{\ell} + 1) (-1)^{1 + J^{\prime} + J_{\ell}^{\prime} + K}
	\sqrt{\frac{(J + J^{\prime} + 1)(2J + 1)(2J^{\prime} + 1)
	(2J_{\ell} + 1)(2J_{\ell}^{\prime} +1)}{J_{\ell} + J_{\ell}^{\prime} 
	+ 1}} \\
	& \times \, 
	\left\{ \begin{array}{ccc}
		J & J_{\ell} & 1 \\
		J_{\ell}^{\prime} & J^{\prime} & K
	\end{array} \right\}
	\left\{ \begin{array}{ccc}
		L & L_{\ell} & 1 \\
		J_{\ell} & J & S
	\end{array} \right\}
	\left\{ \begin{array}{ccc}
		L & L_{\ell} & 1 \\
		J_{\ell}^{\prime} & J^{\prime} & S
	\end{array} \right\}
	\mathcal{C}_I(\beta_{\ell} L_{\ell} S \rightarrow \beta L S) \; .
\label{Eq:CIK_dip}
\end{split}
\end{equation}
This equation can be used to calculate the multipole components of the 
inelastic collision transfer rates for $J$-state interference from the values 
of the usual inelastic collisional rates for $J$-level populations.
As a proof of the consistency of Eq.~(\ref{Eq:CIK_dip}), we observe that the 
0-rank multipole component is given by
\begin{equation}
	C_I^{(0)}(\beta L S J J, \beta_{\ell} L_{\ell} S J_{\ell} J_{\ell}) =
	\mathcal{C}_I(\beta_{\ell} L_{\ell} S J_{\ell} \rightarrow \beta L S J) 
	= (2L_{\ell} +1) (2J + 1) 
	\left\{ \begin{array}{ccc}
		L & L_{\ell} & 1 \\
		J_{\ell} & J & S
	\end{array} \right\}^2
	\mathcal{C}_I(\beta_{\ell} L_{\ell} S \rightarrow \beta L S) \; ,
\label{Eq:CI0p}
\end{equation}
which is the analogous to Eq.~(\ref{Eq:B_rel1}), while using 
Eqs.~(\ref{Eq:6ja}) and (\ref{Eq:CI0p}), the diagonal terms are given by
\begin{equation}
	C_I^{(K)}(\beta L S J J,\beta_{\ell} L_{\ell} S J_{\ell} J_{\ell}) = 
	(-1)^{K} \, \frac{
	\left\{ \begin{array}{ccc}
		J & J & K \\
		J_{\ell} & J_{\ell} & 1
	\end{array} \right\}
	}{
	\left\{ \begin{array}{ccc}
		J & J & 0 \\
		J_{\ell} & J_{\ell} & 1
	\end{array} \right\} }
	\, \mathcal{C}_I(\beta_{\ell} L_{\ell} S J_{\ell} \rightarrow 
	\beta L S J) \; ,
\label{Eq:CIK_diag2}
\end{equation}
which corresponds to Eq.~(\ref{Eq:CIK_diag1}) for $\tilde{K}=1$.

\section{Transfer rate due to superelastic collisions}
A similar reasoning can be followed for the transfer rates due to superelastic 
collisions. 
These transfer processes contribute to the time evolution of a particular 
density matrix element according to the equation
\begin{equation}
	\frac{\rm d}{{\rm d} t} \, 
	\rho_{\beta L S}(J M, J^{\prime}M^{\prime}) =
	\sum_{\beta_u L_u J_u M_u J_u^{\prime} M_u^{\prime}}
	C_S(\beta L S J M J^{\prime} M^{\prime}, \beta_u L_u S J_u M_u
		J_u^{\prime} M_u^{\prime}) \, 
	\rho_{\beta_u L_u S}(J_u M_u, J_u^{\prime} M_u^{\prime}) \; ,
\label{Eq:CS_std}
\end{equation}
where $C_S$ is the superelastic collision transfer rate and where the quantum 
numbers $(\beta_u L_u S)$ denote any term having energy higher than the term 
$(\beta L S)$.
Following the same steps as in Sect.~\ref{Sect:CI}, it can be shown that under 
the assumption of isotropic collisions, the transfer rate $C_S$ can be written 
in the form
\begin{equation}
\begin{split}
	C_S(\beta L S J M J^{\prime} M^{\prime}, \beta_u L_u S J_u M_u
		J_u^{\prime} M_u^{\prime}) = & 
	\sqrt{\frac{J_u + J_u^{\prime} + 1}{J + J^{\prime} + 1}} \,
	(-1)^{J + J_u - M^{\prime} - M_u^{\prime}} \\
	& \times \sum_K (2K + 1) 
	\left( \begin{array}{ccc}
		J^{\prime} & J & K \\
		M^{\prime} & -M & Q 
	\end{array} \right)
	\left( \begin{array}{ccc}
		J_u^{\prime} & J_u & K \\
		M_u^{\prime} & -M_u & Q 
	\end{array} \right)
	C_S^{(K)}(\beta L S J J^{\prime}, \beta_u L_u S J_u J_u^{\prime}) \; ,
\label{Eq:CS_CSK_rel}
\end{split}
\end{equation}
where the multipole components of the superelastic collision transfer rate,
$C_S^{(K)}$, are defined by the equation
\begin{equation}
\begin{split}
	C_S^{(K)}(\beta L S J J^{\prime}, \beta_u L_u S J_u J_u^{\prime}) = &
	\sqrt{\frac{J + J^{\prime} + 1}{J_u + J_u^{\prime} + 1}} \\
	& \times \sum_{N N^{\prime} N_u N_u^{\prime}} 
	(-1)^{J + J_u - N^{\prime} - N_u^{\prime}}
	\left( \begin{array}{ccc}
		J^{\prime} & J & K \\
		N^{\prime} & -N & P 
	\end{array} \right)
	\left( \begin{array}{ccc}
		J_u^{\prime} & J_u & K \\
		N_u^{\prime} & -N_u & P
	\end{array} \right)
	C_S(\beta L S J N J^{\prime} N^{\prime}, \beta_u L_u S J_u N_u
		J_u^{\prime} N_u^{\prime}) \, .
\label{Eq:CSK_def}
\end{split}
\end{equation}
By substituting Eq.~(\ref{Eq:CS_CSK_rel}) into Eq.~(\ref{Eq:CS_std}), and 
recalling Eq.~(\ref{Eq:rho_KQ}), we find the following equation for the 
spherical statistical tensors
\begin{equation}
	\frac{\rm d}{{\rm d} t} \, 
	^{\beta L S} \! \rho^K_Q(J, J^{\prime}) = 
	\sum_{\beta_u L_u J_u J_u^{\prime}}
	\sqrt{\frac{J_u + J_u^{\prime} + 1}{J + J^{\prime} + 1}} \,
	C_S^{(K)}(\beta L S J J^{\prime}, \beta_u L_u S J_u J_u^{\prime}) \;
	^{\beta_u L_u S} \! \rho^K_Q(J_u, J_u^{\prime}) \; .
\end{equation}
The 0-rank multipole component is given by
\begin{equation}
	C_S^{(0)}(\beta L S J J^{\prime}, \beta_u L_u S J_u J_u^{\prime}) =
	\delta_{J J^{\prime}} \, \delta_{J_u J_u^{\prime}} \,
	\frac{1}{2J_u +1} \sum_{N N_u}
	C_S(\beta L S J N J N, \beta_u L_u S J_u N_u J_u N_u) =
	\delta_{J J^{\prime}} \, \delta_{J_u J_u^{\prime}} \,
	\mathcal{C}_S(\beta_u L_u S J_u \rightarrow \beta L S J) \; ,
\label{Eq:CS0_1}
\end{equation}
where $\mathcal{C}_S(\beta_u L_u S J_u \rightarrow \beta L S J)$ is the 
usual superelastic collisional rate for the transition from the upper level 
$|\, \beta_u L_u S J_u \rangle$ to the lower level $|\, \beta L S J \rangle$.
When the interaction between the atomic system and the colliders can be 
described by means of a single operator of rank $\tilde{K}$, it can be shown 
that the multipole components of rank $K$ of the diagonal rates ($J=J^{\prime}$ 
and $J_u = J_u^{\prime}$) are related to the multipole components of rank 0 by 
the equation
\begin{equation}
	C_S^{(K)}(\beta L S J J, \beta_u L_u S J_u J_u) =
	(-1)^{K} \frac{
	\left\{ \begin{array}{ccc}
		J & J & K \\
		J_u & J_u & \tilde{K}
	\end{array} \right\}
	}{
	\left\{ \begin{array}{ccc}
		J & J & 0 \\
		J_u & J_u & \tilde{K}
	\end{array} \right\} }
	C_S^{(0)}(\beta L S J J, \beta_u L_u S J_u J_u) =
	(-1)^{K} \frac{
	\left\{ \begin{array}{ccc}
		J & J & K \\
		J_u & J_u & \tilde{K}
	\end{array} \right\}
	}{
	\left\{ \begin{array}{ccc}
		J & J & 0 \\
		J_u & J_u & \tilde{K}
	\end{array} \right\} }
	\mathcal{C}_S(\beta_u L_u S J_u \rightarrow \beta L S J) \; .
\label{Eq:CSK_diag1}
\end{equation}
As discussed in the previous section for the case of inelastic collisions, a 
similar relation for the nondiagonal rates (describing the transfer of 
$J$-state interference due to superelastic collisions) can be derived under 
the assumption that the electron-atom interaction is described by a dipolar 
operator.
By analogy with the expression of the multiterm atom radiative transfer rate 
due to stimulated emission processes ($\mathbb{T}_S$, see Eq.~(7.45c) of LL04) 
in the presence of an isotropic incident field, we find the following relation
\begin{equation}
\begin{split}
	C_S^{(K)}(\beta L S J J^{\prime},\beta_u L_u S J_u J_u^{\prime}) 
	= & (2L_u + 1) (-1)^{1 + J^{\prime} + J_u^{\prime} + K}
	\sqrt{\frac{(J + J^{\prime} + 1)(2J + 1)(2J^{\prime} + 1)
	(2J_u + 1)(2J_u^{\prime} +1)}{J_u + J_u^{\prime} + 1}} \\
	& \times \, 
	\left\{ \begin{array}{ccc}
		J & J_u & 1 \\
		J_u^{\prime} & J^{\prime} & K
	\end{array} \right\}
	\left\{ \begin{array}{ccc}
		L & L_u & 1 \\
		J_u & J & S
	\end{array} \right\}
	\left\{ \begin{array}{ccc}
		L & L_u & 1 \\
		J_u^{\prime} & J^{\prime} & S
	\end{array} \right\}
	\mathcal{C}_S(\beta_u L_u S \rightarrow \beta L S) \; ,
\label{Eq:CSK_dip}
\end{split}
\end{equation}
where we have introduced the superelastic collisional rate for the transition 
from the upper to the lower term $\mathcal{C}_S(\beta_u L_u S \rightarrow 
\beta L S)$, defined by
\begin{equation}
	\mathcal{C}_S(\beta_u L_u S \rightarrow \beta L S) =
	\sum_{J} \mathcal{C}_S(\beta_u L_u S J_u \rightarrow \beta L S J) \; ,
\label{Eq:CS_term1}
\end{equation}
the sum being extended to all the $J$-levels of the lower term to which a 
given $J$-level of the upper term can be connected through an electric dipole 
transition.\footnote{The sum appearing on the righthand side of 
Eq.~(\ref{Eq:CS_term1}) does not depend on the particular $J$-level of the 
upper term that is considered.}

\section{Relaxation rates due to inelastic and superelastic collisions}
In a given reference system, relaxation processes due to inelastic and 
superelastic collisions contribute to the time evolution of a particular 
density-matrix element via an equation of the form
\begin{equation}
	\frac{\rm d}{{\rm d} t} \, 
	\rho_{\beta L S}(J M, J^{\prime}M^{\prime}) =
	- \sum_{J^{\prime \prime} M^{\prime \prime}}
	\left[ f(\beta L S J M J^{\prime} M^{\prime} J^{\prime \prime} 
		M^{\prime \prime}) \,
	\rho_{\beta L S}(J M, J^{\prime \prime} M^{\prime \prime}) +
	g(\beta L S J M J^{\prime} M^{\prime} J^{\prime \prime} 
		M^{\prime \prime}) \,
	\rho_{\beta L S}(J^{\prime \prime} M^{\prime \prime}, J^{\prime} 
		M^{\prime}) \right] \; .
\label{Eq:RLX_std1}
\end{equation}
The conjugation property of the density-matrix elements 
($\rho_{\beta L S}(J M, J^{\prime} M^{\prime})^{\ast} = 
\rho_{\beta L S}(J^{\prime} M^{\prime}, J M)$) requires that
\begin{equation}
	g(\beta L S J M J^{\prime} M^{\prime} J^{\prime \prime} 
		M^{\prime \prime}) = 
	f(\beta L S J^{\prime} M^{\prime} J M J^{\prime \prime} 
		M^{\prime \prime})^{\ast} \; ,
\end{equation}
so that Eq.~(\ref{Eq:RLX_std1}) can be written in the form
\begin{equation}
	\frac{\rm d}{{\rm d} t} \, 
	\rho_{\beta L S}(J M, J^{\prime}M^{\prime}) =
	- \sum_{J^{\prime \prime} M^{\prime \prime}}
	\left[ \frac{1}{2} S(\beta L S J M J^{\prime} M^{\prime} 
		J^{\prime \prime} M^{\prime \prime}) \,
	\rho_{\beta L S}(J M, J^{\prime \prime} M^{\prime \prime}) +
	\frac{1}{2} S(\beta L S J^{\prime} M^{\prime} J M J^{\prime \prime} 
		M^{\prime \prime})^{\ast} \,
	\rho_{\beta L S}(J^{\prime \prime} M^{\prime \prime}, J^{\prime} 
		M^{\prime}) \right] \; .
\label{Eq:RLX_std2}
\end{equation}

In a new reference system, obtained from the old one by the rotation $R$, 
recalling Eqs.~(\ref{Eq:rot1}) and (\ref{Eq:rot2}), we have
\begin{equation}
\begin{split}
	\frac{\rm d}{{\rm d} t} 
	\left[ \rho_{\beta L S}(J M, J^{\prime} M^{\prime}) \right]_{\rm new} =
	-\sum_{J^{\prime \prime} M^{\prime \prime} M^{\prime \prime \prime}}
	\Bigg\{ & \frac{1}{2} \sum_{N N^{\prime} N^{\prime \prime}}
	{\mathcal D}^{J}_{N M}(R)^{\ast} \,
	{\mathcal D}^{J^{\prime}}_{N^{\prime} M^{\prime}}(R) \,
	S(\beta L S J N J^{\prime} N^{\prime} J^{\prime \prime} 
		N^{\prime \prime}) \\
	& \phantom{\Bigg\{ } \quad \quad \quad \times \, 
	{\mathcal D}^{J}_{N M^{\prime \prime \prime}}(R) \,
	{\mathcal D}^{J^{\prime \prime}}_{N^{\prime \prime} M^{\prime \prime}}
		(R)^{\ast} \,
	\left[ \rho_{\beta L S}(J M^{\prime \prime \prime}, 
		J^{\prime \prime} M^{\prime \prime}) \right]_{\rm new} \\ 
	& \phantom{\Bigg\{ } \!\!\!\!\!\!\!\!\!\! + \, 
	\frac{1}{2} \sum_{N N^{\prime} N^{\prime \prime}} 
	{\mathcal D}^{J}_{N M}(R)^{\ast} \,
	{\mathcal D}^{J^{\prime}}_{N^{\prime} M^{\prime}}(R) \,
	S(\beta L S J^{\prime} N^{\prime} J N J^{\prime \prime} 
		N^{\prime \prime})^{\ast} \\
	& \phantom{\Bigg\{ } \quad \quad \quad \times \, 
	{\mathcal D}^{J^{\prime \prime}}_{N^{\prime \prime} M^{\prime \prime}}
		(R) \,
	{\mathcal D}^{J^{\prime}}_{N^{\prime} M^{\prime \prime \prime}}
		(R)^{\ast} \,
	\left[ \rho_{\beta L S}(J^{\prime \prime} M^{\prime \prime}, 
		J^{\prime} M^{\prime \prime \prime}) \right]_{\rm new} 
	\Bigg\} \, .
\label{Eq:RLX_std2_new}
\end{split}
\end{equation}
Due to the isotropy of collisions, Eqs.~(\ref{Eq:RLX_std2}) and 
(\ref{Eq:RLX_std2_new}) must be identical, which implies
\begin{equation}
	S(\beta L S J M J^{\prime} M^{\prime} J^{\prime \prime} 
		M^{\prime \prime})
	\, \delta_{M M^{\prime \prime \prime}} = 
	\sum_{N N^{\prime} N^{\prime \prime}} 
	{\mathcal D}^{J}_{N M}(R)^{\ast} \,
	{\mathcal D}^{J^{\prime}}_{N^{\prime} M^{\prime}}(R) \,
	S(\beta L S J N J^{\prime} N^{\prime} J^{\prime \prime} 
		N^{\prime \prime}) \,
	{\mathcal D}^{J}_{N M^{\prime \prime \prime}}(R) \,
	{\mathcal D}^{J^{\prime \prime}}_{N^{\prime \prime} M^{\prime \prime}}
		(R)^{\ast} \; ,
\label{Eq:RLX_rel1}
\end{equation}
regardless of the rotation $R$.
This requires the rate $S(\beta L S J N J^{\prime} N^{\prime} J^{\prime \prime} 
N^{\prime \prime})$ to be independent of the quantum number $N$ (if not, 
the righthand side of Eq.~(\ref{Eq:RLX_rel1}) would not be zero for 
$M \ne M^{\prime \prime \prime}$, no matter the rotation $R$).
We can thus carry out the summation over $N$ via Eq.~(\ref{Eq:rotmat_orto}) to 
get (with the help of Eq.~(\ref{Eq:rotmat1}))
\begin{equation}
	S(\beta L S J J^{\prime} M^{\prime} J^{\prime \prime} M^{\prime \prime})
	= \sum_{N^{\prime} N^{\prime \prime}} 
	S(\beta L S J J^{\prime} N^{\prime} J^{\prime \prime} N^{\prime \prime})
	\, (-1)^{N^{\prime \prime} - M^{\prime \prime}}
	\sum_K (2K + 1) 
	\left( \begin{array}{ccc}
		J^{\prime} & J^{\prime \prime} & K \\
		N^{\prime} & -N^{\prime \prime} & P
	\end{array} \right)
	\left( \begin{array}{ccc}
		J^{\prime} & J^{\prime \prime} & K \\
		M^{\prime} & -M^{\prime \prime} & Q
	\end{array} \right)
	\mathcal{D}^K_{P Q}(R)^{\ast} \; .
\label{Eq:RLX_rel2}
\end{equation}
Since the righthand side of Eq.~(\ref{Eq:RLX_rel2}) must be independent 
of the rotation $R$, index $K$ can only take the value $K=0$, which implies 
$Q=P=0$, $N^{\prime}=N^{\prime \prime}$, $M^{\prime} = M^{\prime \prime}$, 
and $J^{\prime} = J^{\prime \prime}$ from Eq.~(\ref{Eq:3jc}).
We thus obtain
\begin{equation}
	S(\beta L S J J^{\prime} M^{\prime} J^{\prime \prime} M^{\prime \prime})
	= \delta_{M^{\prime} M^{\prime \prime}} 
	\delta_{J^{\prime} J^{\prime \prime}} 
	\frac{1}{2J^{\prime} + 1} \sum_{N^{\prime}} 
	S(\beta L S J J^{\prime} N^{\prime} J^{\prime} N^{\prime}) \; .
\end{equation}
Substitution into Eq.~(\ref{Eq:RLX_std2}) gives
\begin{equation}
	\frac{\rm d}{{\rm d} t} \, 
	\rho_{\beta L S}(J M, J^{\prime}M^{\prime}) =
	- S_0(\beta L S J J^{\prime}) \, 
	\rho_{\beta L S}(J M, J^{\prime} M^{\prime}) \; ,
\label{Eq:RLX_std3}
\end{equation}
or, in the spherical statistical tensor representation,
\begin{equation}
	\frac{\rm d}{{\rm d} t} \, 
	^{\beta L S} \! \rho^K_Q(J, J^{\prime}) = 
	- S_0(\beta L S J J^{\prime}) \, 
	^{\beta L S} \! \rho^K_Q(J, J^{\prime}) \; , 
\label{Eq:RLX_std4}
\end{equation}
where we have introduced the collisional relaxation rate
\begin{equation}
	S_0(\beta L S J J^{\prime}) = \frac{1}{2} \left[ 
	\frac{1}{2J^{\prime} + 1} \sum_{M^{\prime}} 
	S(\beta L S J J^{\prime} M^{\prime} J^{\prime} M^{\prime}) + 
	\frac{1}{2J + 1} \sum_M  S(\beta L S J^{\prime} J M J M)^{\ast} \right] 
	\; .
\end{equation}
The diagonal element
\begin{equation}
	S_0(\beta L S J J) = \frac{1}{2J + 1} \frac{1}{2}
	\sum_M \left[ S(\beta L S J J M J M) + S(\beta L S J J M J M)^{\ast} 
	\right] = \frac{1}{2J +1} {\rm Re} \left[ \sum_M 
	S(\beta L S J J M J M)) \right]
\end{equation}
coincides with the one defined in LL04 for the case of a multilevel atom.

As shown in LL04, the diagonal element $S_0(\beta L S J J)$, which represents 
the relaxation rate of populations and interference between magnetic sublevels 
pertaining to the same $J$-level (see Eqs.~(\ref{Eq:RLX_std3}) and 
(\ref{Eq:RLX_std4})), is connected to the 0-rank multipole components of the 
inelastic and superelastic collision transfer rates by the equation
\begin{equation}
	S_0(\beta L S J J) = \sum_{\beta_u L_u J_u} 
	C_I^{(0)}(\beta_u L_u S J_u J_u, \beta L S J J) + 
	\sum_{\beta_{\ell} L_{\ell} J_{\ell}}
	C_S^{(0)}(\beta_{\ell} L_{\ell} S J_{\ell} J_{\ell}, \beta L S J J) \; .
\label{Eq:S0JJ}
\end{equation}

To obtain a similar relation for the nondiagonal elements 
($S_0(\beta L S J J^{\prime})$ with $J \ne J^{\prime}$), which represent the 
relaxation rate of $J$-state interference due to inelastic and superelastic 
collisions, we make the assumption, also in this case, that the interaction 
between the atoms and colliders is described by a dipolar operator, and we 
proceed by analogy with the multiterm atom radiative relaxation rates due to 
absorption ($\mathbb{R}_A$) and stimulated emission ($\mathbb{R}_S$) processes 
(see Eqs.~(7.46a) and (7.46c) of LL04).
Assuming an isotropic incident radiation field (i.e. setting $K_r=0$), such
radiative rates assume the simple form
\begin{align}
	\mathbb{R_A}(\beta L S K Q J J^{\prime} K Q J J^{\prime}) = &
	\sum_{\beta_u L_u} B(\beta L S \rightarrow \beta_u L_u S) J^0_0 \; , \\
	\mathbb{R_S}(\beta L S K Q J J^{\prime} K Q J J^{\prime}) = &
	\sum_{\beta_{\ell} L_{\ell}} 
	B(\beta L S \rightarrow \beta_{\ell} L_{\ell} S) J^0_0 \; .
\end{align}
Introducing the inelastic and superelastic collisional rates for transitions 
between different terms (see Eqs.~(\ref{Eq:CI_term1}) and (\ref{Eq:CS_term1})), 
we have
\begin{align}
	S_0(\beta L S J J^{\prime}) = &
	\sum_{\beta_u L_u} \mathcal{C}_I(\beta L S \rightarrow \beta_u L_u S) +
	\sum_{\beta_{\ell} L_{\ell}} 
	\mathcal{C}_S(\beta L S \rightarrow \beta_{\ell} L_{\ell} S) 
	\nonumber \\
	= & \sum_{\beta_u L_u J_u} 
	\mathcal{C}_I(\beta L S J \rightarrow \beta_u L_u S J_u) + 
	\sum_{\beta_{\ell} L_{\ell} J_{\ell}} 
	\mathcal{C}_S(\beta L S J \rightarrow \beta_{\ell} L_{\ell} S 
	J_{\ell}) = S_0(\beta L S J J) = S_0(\beta L S) \; .
\label{Eq:S0JJp}
\end{align}
The relaxation rate of $J$-state interference due to inelastic and superelastic 
collisions thus coincides with the relaxation rate of $J$-level populations 
and of interference between magnetic sublevels pertaining to the same 
$J$-level. 
This rate, on the other hand, does not depend on the quantum number $J$, and 
is thus identical for all the $J$-levels of a given term.

When collected together transfer and relaxation rates, the statistical 
equilibrium equations for the spherical statistical tensors can be written in 
the form
\begin{equation}
\begin{split}
	\frac{\rm d}{{\rm d} t} \, 
	^{\beta L S} \! \rho^K_Q(J, J^{\prime}) = &
	\sum_{\beta_{\ell} L_{\ell} J_{\ell} J_{\ell}^{\prime}}
	\sqrt{\frac{J_{\ell} + J_{\ell}^{\prime} + 1}{J + J^{\prime} + 1}} \,
	C_I^{(K)}(\beta L S J J^{\prime}, \beta_{\ell} L_{\ell} S J_{\ell} 
		J_{\ell}^{\prime}) \;
	^{\beta_{\ell} L_{\ell} S} \! \rho^K_Q(J_{\ell}, J_{\ell}^{\prime}) \\
	& + \sum_{\beta_u L_u J_u J_u^{\prime}}
	\sqrt{\frac{J_u + J_u^{\prime} + 1}{J + J^{\prime} + 1}} \,
	C_S^{(K)}(\beta L S J J^{\prime}, \beta_u L_u S J_u J_u^{\prime}) \;
	^{\beta_u L_u S} \! \rho^K_Q(J_u, J_u^{\prime}) \\
	& - S_0(\beta L S J J^{\prime}) \,
	{^{\beta L S} \! \rho^K_Q}(J, J^{\prime}) \; .
\end{split}
\end{equation}

\section{Application to the case of a two-term atom with unpolarized lower term}
We consider a two-term atom and denote the quantum numbers characterizing the 
lower and upper term by $(\beta_{\ell} L_{\ell} S)$ and $(\beta_u L_u S)$, 
respectively.
The time evolution of the spherical statistical tensors of the upper term, 
when taking both radiative (see Eq.~(10.115) of LL04) and collisional 
(inelastic and superelastic collisions only) processes into account is 
described by the equation
\begin{equation}
\begin{split}
	\frac{\rm d}{{\rm d} t} \, 
	^{\beta_u L_u S} \! \rho^K_Q(J_u, J_u^{\prime}) = &
	- 2 \pi {\rm i} \sum_{K^{\prime} Q^{\prime} J_u^{\prime \prime} 
	J_u^{\prime \prime \prime}} N_{\beta_u L_u S}(K Q J_u J_u^{\prime}, 
	K^{\prime} Q^{\prime} J_u^{\prime \prime} J_u^{\prime \prime \prime})
	\; ^{\beta_u L_u S} \! \rho^{K^{\prime}}_{Q^{\prime}}
		(J_u^{\prime \prime}, J_u^{\prime \prime \prime}) \\
	& + \sum_{K^{\prime} Q^{\prime} J_{\ell} J_{\ell}^{\prime}}
	\mathbb{T}_A(\beta_u L_u S K Q J_u J_u^{\prime}, \beta_{\ell} L_{\ell}
		S K^{\prime} Q^{\prime} J_{\ell} J_{\ell}^{\prime}) \;
	^{\beta_{\ell} L_{\ell} S} \! \rho^{K^{\prime}}_{Q^{\prime}}
		(J_{\ell}, J_{\ell}^{\prime}) \\
	& - \sum_{K^{\prime} Q^{\prime} J_u^{\prime \prime} 
		J_u^{\prime \prime \prime}}
	\bigg[ \mathbb{R}_E(\beta_u L_u S K Q J_u J_u^{\prime} K^{\prime}
		Q^{\prime} J_u^{\prime \prime} J_u^{\prime \prime \prime}) \\
	& \qquad \qquad \;\;\; +
	\mathbb{R}_S(\beta_u L_u S K Q J_u J_u^{\prime} K^{\prime}
		Q^{\prime} J_u^{\prime \prime} J_u^{\prime \prime \prime}) 
	\bigg] \;
	^{\beta_u L_u S} \! \rho^{K^{\prime}}_{Q^{\prime}}
		(J_u^{\prime \prime}, J_u^{\prime \prime \prime}) \\
	& + \sum_{J_{\ell} J_{\ell}^{\prime}} 
	\sqrt{\frac{J_{\ell} +J_{\ell}^{\prime} +1}{J_u +J_u^{\prime} +1}} \,
	C_I^{(K)}(\beta_u L_u S J_u J_u^{\prime}, \beta_{\ell} L_{\ell} S 
		J_{\ell} J_{\ell}^{\prime}) \;
	^{\beta_{\ell} L_{\ell} S} \! \rho^{K}_{Q}
		(J_{\ell}, J_{\ell}^{\prime}) \\
	& - S_0(\beta_u L_u S J_u J_u^{\prime}) \;
	^{\beta_u L_u S} \! \rho^K_Q(J_u, J_u^{\prime}) \; ,
\end{split}
\end{equation}
where $N$ is the magnetic kernel (see Eq.~(7.41) of LL04), $\mathbb{T}_A$ the 
radiative transfer rate due to absorption, while $\mathbb{R}_E$ and 
$\mathbb{R}_S$ are the radiative relaxation rates due to spontaneous and 
stimulated emission, respectively.

We now make the following simplifying assumptions:
\begin{itemize}
	\item{There is no magnetic field. Under this assumption the kernel $N$ 
		takes the simpler form
		\begin{equation}
			N_{\beta_u L_u S}(K Q J_u J_u^{\prime}, K^{\prime} 
				Q^{\prime} J_u^{\prime \prime} 
				J_u^{\prime \prime \prime}) =
			\delta_{K K^{\prime}} \, \delta_{Q Q^{\prime}} \,
			\delta_{J_u J_u^{\prime \prime}} \,
			\delta_{J_u^{\prime} J_u^{\prime \prime \prime}} \,
			\nu_{\beta_u L_u S J_u, \, \beta_u L_u S 
				J_u^{\prime}} \; ,
		\end{equation}
		with $\nu_{\beta_u L_u S J_u, \, \beta_u L_u S	J_u^{\prime}}=
		[E(\beta_u L_u S J_u) - E(\beta_u L_u S J_u^{\prime})]/h$, 
		where $E(\beta L S J)$ is the energy of a given fine-structure 
		$J$-level, and $h$ is the Planck constant.}
	\item{The radiation field is weak so that stimulated emission can be 
		neglected ($\mathbb{R}_S=0$).}
	\item{The lower term is unpolarized (i.e., the magnetic sublevels of 
		the lower term are evenly populated and no interference is 
		present between them).
		Under this assumption the spherical statistical tensors of the 
		lower term are given by 
		\begin{equation}
			^{\beta_{\ell} L_{\ell} S} \! \rho^{K}_{Q}
				(J_{\ell}, J_{\ell}^{\prime}) =
			\delta_{K 0} \, \delta_{Q 0} \,
			\delta_{J_{\ell} J_{\ell}^{\prime}} \,
			\frac{\sqrt{2J_{\ell} + 1}}{(2S +1)(2L_{\ell} + 1)}
			\frac{\mathbb{N}_{\ell}}{\mathcal{N}} \; ,
		\end{equation}
		where $\mathcal{N}$ is total number density of atoms, and 
		$\mathbb{N}_{\ell}$ the number density of atoms in the 
		lower term.}
	\item{The electron-atom interaction is described by a 
		dipolar operator.
		Under this assumption, defining through Eq.~(\ref{Eq:CS_term1}) 
		a superelastic collisional rate for the transition from the 
		upper to the lower term ($\mathcal{C}_S(\beta_u L_u S 
		\rightarrow \beta_{\ell} L_{\ell} S)$), the collisional 
		relaxation rate is given by (see Eq.~(\ref{Eq:S0JJp}))
		\begin{equation}
			S_0(\beta_u L_u S J_u J_u^{\prime}) = 
			\mathcal{C}_S(\beta_u L_u S \rightarrow \beta_{\ell} 
			L_{\ell} S) \; .
		\end{equation}}
\end{itemize}
Taking the above-mentioned assumptions into account, and recalling that 
(see Eq.~(7.46b) of LL04) $\mathbb{R}_E(\beta_u L_u S K Q J_u J_u^{\prime} 
K^{\prime} Q^{\prime} J_u^{\prime \prime} J_u^{\prime \prime \prime}) = 
\delta_{K K^{\prime}} \, \delta_{Q Q^{\prime}} \, 
\delta_{J_u J_u^{\prime \prime}} \, 
\delta_{J_u^{\prime} J_u^{\prime \prime \prime}}
A(\beta_u L_u S \rightarrow \beta_{\ell} L_{\ell} S)$, we obtain
\begin{equation}
\begin{split}
	\frac{\rm d}{{\rm d} t} \, 
	^{\beta_u L_u S} \! \rho^K_Q(J_u, J_u^{\prime}) = &
	- 2 \pi {\rm i} 
	\nu_{\beta_u L_u S J_u, \, \beta_u L_u S J_u^{\prime}} 
	\; ^{\beta_u L_u S} \! \rho^{K}_{Q}(J_u, J_u^{\prime}) \\
	& + \sum_{J_{\ell}}
	\mathbb{T}_A(\beta_u L_u S K Q J_u J_u^{\prime}, \beta_{\ell} L_{\ell}
		S 0 0 J_{\ell} J_{\ell}) \;
	\frac{\sqrt{2J_{\ell} + 1}}{(2S +1)(2L_{\ell} + 1)}
	\frac{\mathbb{N}_{\ell}}{\mathcal{N}} \\
	& - A(\beta_u L_u S \rightarrow \beta_{\ell} L_{\ell} S) \; 
	^{\beta_u L_u S} \! \rho^{K}_{Q}(J_u, J_u^{\prime}) \\
	& + \delta_{K 0} \, \delta_{Q 0} \sum_{J_{\ell}} 
	\sqrt{\frac{2J_{\ell} + 1}{2J_u + 1}} \,
	C_I^{(0)}(\beta_u L_u S J_u J_u, \beta_{\ell} L_{\ell} S 
		J_{\ell} J_{\ell})
	\frac{\sqrt{2J_{\ell} + 1}}{(2S +1)(2L_{\ell} + 1)}
	\frac{\mathbb{N}_{\ell}}{\mathcal{N}} \\
	& - \mathcal{C}_S(\beta_u L_u S \rightarrow \beta_{\ell} L_{\ell} S) 
	\; ^{\beta_u L_u S} \! \rho^K_Q(J_u, J_u^{\prime}) \; .
	\label{Eq:SEE_2}
\end{split}
\end{equation}
As expected, under the hypotheses of isotropic collisions and unpolarized lower 
term, transfer processes due to inelastic collisions only contribute to the 
time evolution of the 0-rank spherical statistical tensors of the upper term.
Assuming that the colliding particles are characterized by a Maxwellian 
velocity distribution, the collisional rates $C_S^{(0)}$ and $C_I^{(0)}$ can be 
related through the Milne-Einstein relation 
\begin{equation}
	C_S^{(0)}(\beta_{\ell} L_{\ell} S J_{\ell} J_{\ell}, \beta_u L_u S 
		J_u J_u) = \frac{2J_{\ell} + 1}{2J_u +1} 
	C_I^{(0)}(\beta_u L_u S J_u J_u, \beta_{\ell} L_{\ell} S J_{\ell} 
		J_{\ell}) \,
	{\rm exp} \left[ \frac{E(\beta_u L_u S J_u) - E(\beta_{\ell} L_{\ell} 
		S J_{\ell})}{K_B T} \right] \; ,
\end{equation}
where $T$ is the electron temperature.
Using Eq.~(\ref{Eq:CS_term1}), the fourth term on the righthand side of 
Eq.~(\ref{Eq:SEE_2}) can be written in the form
\begin{equation}
	\delta_{K 0} \, \delta_{Q 0} \,
	\mathcal{C}_S(\beta_u L_u S \rightarrow \beta_{\ell} L_{\ell} S) \,
	\frac{c^2}{2 h \nu_0^3} \, B_T(\nu_0) \,
	\frac{\sqrt{2J_u + 1}}{(2S +1)(2L_{\ell} + 1)}
	\frac{\mathbb{N}_{\ell}}{\mathcal{N}} \; ,
\end{equation}
where $B_T(\nu_0)$ is the Planck function in the Wien limit (consistently with 
the assumption of neglecting stimulated emission) at temperature $T$, and 
where $\nu_0$ is the Bohr frequency corresponding to the energy difference 
between the centers of gravity of the two terms. (We neglect the frequency 
differences among the various components of the multiplet in the exponential 
appearing in the Milne-Einstein relation.)

Taking the expression of $\mathbb{T}_A(\beta_u L_u S K Q J_u J_u^{\prime}, 
\beta_{\ell} L_{\ell} S 0 0 J_{\ell} J_{\ell})$ (see Eq.~(10.124) of LL04) into 
account and performing the sum over $J_{\ell}$ using Eq.~(\ref{Eq:6j_sum2}), 
the second term on the righthand side of Eq.~(\ref{Eq:SEE_2}) is given 
by
\begin{equation}
	B(\beta_{\ell} L_{\ell} S \rightarrow \beta_u L_u S) \,
	(-1)^{1 - L_{\ell} + S + J_u^{\prime} + K + Q}
	\left\{ \begin{array}{ccc}
		1 & 1 & K \\
		L_u & L_u & L_{\ell} 
	\end{array} \right\}
	\left\{ \begin{array}{ccc}
		L_u & L_u & K \\
		J_u & J_u^{\prime} & S
	\end{array} \right\}
	J^K_{-Q}(\nu_0) \,
	\frac{\sqrt{3(2J_u + 1)(2J_u^{\prime} + 1)}}{2S+1}
	\frac{\mathbb{N}_{\ell}}{\mathcal{N}} \; .
\end{equation}
We recall that the quantum theory of polarization described in LL04 is valid 
under the so-called flat spectrum approximation (that is, the incident 
radiation field that produces optical pumping in the atomic system must be 
flat over a frequency interval $\Delta{\nu}$ larger than the natural width of 
the atomic levels, and, when coherence between nondegenerate levels is 
involved, $\Delta{\nu}$ must then be larger than the corresponding Bohr 
frequency).
For this reason, it is sufficient to evaluate the radiation field tensor 
$J^K_Q$ (see Eq.~(5.157) of LL04 for its definition) at a single frequency
within the multiplet.

In stationary situations, recalling the relations among the Einstein 
coefficients
\begin{equation}
	B(\beta_{\ell} L_{\ell} S \rightarrow \beta_u L_u S) = 
	\frac{2L_u +1}{2L_{\ell} +1} 
	B(\beta_u L_u S \rightarrow \beta_{\ell} L_{\ell} S) = 
	\frac{2L_u +1}{2L_{\ell} +1} 
	\frac{c^2}{2 h \nu_0^3} \,
	A(\beta_u L_u S \rightarrow \beta_{\ell} L_{\ell} S) \; , 
\end{equation}
the spherical statistical tensors of the upper term are given by
\begin{equation}
\begin{split}
	^{\beta_u L_u S} \! \rho^K_Q(J_u, J_u^{\prime}) = &
	\frac{\sqrt{2J_u +1}}{(2S+1)(2L_u+1)} 
	\frac{\mathbb{N}_{\ell}}{\mathcal{N}}
	\frac{B(\beta_{\ell} L_{\ell} S \rightarrow \beta_u L_u S)}{A(\beta_u 
	L_u S \rightarrow \beta_{\ell} L_{\ell} S)} \\
	& \times \, 
	\frac{(2L_u + 1) \sqrt{3 (2J_u^{\prime} + 1)} \,
	(-1)^{1 - L_{\ell} + S + J_u^{\prime} + K + Q}
	\left\{ \begin{array}{ccc}
		1 & 1 & K \\
		L_u & L_u & L_{\ell} 
	\end{array} \right\}
	\left\{ \begin{array}{ccc}
		L_u & L_u & K \\
		J_u & J_u^{\prime} & S
	\end{array} \right\}
	J^K_{-Q}(\nu_0) +
	\epsilon^{\prime} \, B_T(\nu_0) \, \delta_{K 0} \, \delta_{Q 0}}
	{1 + \epsilon^{\prime} + 2 \pi {\rm i} \nu_{\beta_u L_u S J_u, \, 
		\beta_u L_u S J_u^{\prime}} /
	A(\beta_u L_u S \rightarrow \beta_{\ell} L_{\ell} S)} \; ,
\label{Eq:rhoKQ_col}
\end{split}
\end{equation}
where, in analogy with the two-level atom case, we have introduced the 
quantity 
\begin{equation}
	\epsilon^{\prime} = \frac{\mathcal{C}_S(\beta_u L_u S \rightarrow 
	\beta_{\ell} L_{\ell} S)}{A(\beta_u L_u S \rightarrow \beta_{\ell} 
	L_{\ell} S)} \; .
\end{equation}
It can be easily proven that if $S=0$, so that the upper and lower terms 
are composed by a single fine-structure $J$-level, the expression of a 
two-level atom is recovered (see Eq.~(10.50) of LL04, with 
$H_u=\delta_u^{(K)}=0$).

Substituting Eq.~(\ref{Eq:rhoKQ_col}) into Eq.~(10.127) of LL04, and 
introducing the frequency-integrated absorption coefficient of the multiplet 
\begin{equation}
	k_M^A = \frac{h \nu_0}{4 \pi} \mathbb{N}_{\ell} 
	B(\beta_{\ell} L_{\ell} S \rightarrow \beta_u L_u S) \; ,
\end{equation}
and the absorption profile of the multiplet (in the absence of magnetic fields, 
and for the case of a two-term atom with unpolarized lower term)
\begin{equation}
	\varphi(\nu) = \sum_{J_u J_{\ell}} 
	\frac{(2J_{\ell} + 1)(2J_u +1)}{2S+1} 
	\left\{ \begin{array}{ccc}
		L_u & L_{\ell} & 1 \\
		J_{\ell} & J_u & S 
	\end{array} \right\}^2 
	\phi(\nu_{\beta_u L_u S J_u, \, \beta_{\ell} L_{\ell} S J_{\ell}} -\nu)
	\; ,
\label{Eq:gen_prof}
\end{equation}
where $\phi(\nu_0-\nu)$ are Lorentzian profiles centered at the frequencies of 
the various components of the multiplet, we find the following expression of 
the emission coefficient in the four Stokes parameters:
\begin{equation}
\begin{split}
	\varepsilon_i(\nu,\vec{\Omega}) = & k_M^A \frac{2L_u +1}{2S+1}
	\sum_{KQ} \sum_{J_u J_u^{\prime} J_{\ell}}
	(-1)^{S - L_{\ell} + J_u + J_u^{\prime} + J_{\ell} + K + Q}
	\, 3 (2J_u +1) (2J_u^{\prime} +1) (2J_{\ell} +1) \\
	& \times \, 
	\left\{ \begin{array}{ccc}
		L_u & L_{\ell} & 1 \\
		J_{\ell} & J_u & S 
	\end{array} \right\}
	\left\{ \begin{array}{ccc}
		L_u & L_{\ell} & 1 \\
		J_{\ell} & J_u^{\prime} & S 
	\end{array} \right\}
	\left\{ \begin{array}{ccc}
		1 & 1 & K \\
		J_u & J_u^{\prime} & J_{\ell}
	\end{array} \right\}
	\left\{ \begin{array}{ccc}
		1 & 1 & K \\
		L_u & L_u & L_{\ell}
	\end{array} \right\}
	\left\{ \begin{array}{ccc}
		L_u & L_u & K \\
		J_u & J_u^{\prime} & S
	\end{array} \right\} \\
	& \times \, \mathcal{T}^K_Q(i,\vec{\Omega}) \, 
	J^K_{-Q}(\nu_0) \, \frac{1}{2} \,
	\frac{\Phi(\nu_{\beta_u L_u S J_u, \, \beta_{\ell} L_{\ell} S 
		J_{\ell}} -\nu) + 
	\Phi(\nu_{\beta_u L_u S J_u^{\prime}, \, \beta_{\ell} L_{\ell} S 
		J_{\ell}} -\nu)^{\ast}}
	{1 + \epsilon^{\prime} + 2 \pi {\rm i} \nu_{\beta_u L_u S J_u^{\prime},
		\, \beta_u L_u S J_u} / 
	A(\beta_u L_u S \rightarrow \beta_{\ell} L_{\ell} S)} \\
	& + \, \frac{\epsilon^{\prime}}{1+\epsilon^{\prime}} \, k_M^A \, 
	B_T(\nu_0) \, \varphi(\nu) \, \delta_{i,0} \; ,
\label{Eq:emission1}
\end{split}
\end{equation}
with $i=0$, 1, 2, and 3, standing for Stokes $I$, $Q$, $U$ and $V$, 
respectively. 
Here, $\nu$ and $\vec{\Omega}$ are the frequency and propagation direction of 
the emitted radiation, respectively, $\mathcal{T}^K_Q(i,\vec{\Omega})$ is the 
geometrical tensor introduced by \citet{Lan83}, and $\Phi(\nu_0 - \nu)$ is the 
complex emission profile 
\begin{equation}
	\Phi(\nu_0- \nu) = \phi(\nu_0- \nu) + {\rm i} \, \psi(\nu_0- \nu) \; , 
\end{equation}
with $\phi(\nu_0 - \nu)$ a Lorentzian profile and $\psi(\nu_0 - \nu)$ the 
associated dispersion profile.\footnote{The equations derived here are 
valid in the atom rest frame. Nevertheless, under the assumption of complete 
redistribution on velocities (see Chapter 13 of LL04), the same equations can  
also be applied in the observer's frame, with $\phi(\nu_0 - \nu)$ and 
$\psi(\nu_0 - \nu)$ the Voigt profile and the Faraday-Voigt profile, 
respectively (provided that the atoms have a Maxwellian distribution of 
velocities).}
The last term on the righthand side of Eq.~(\ref{Eq:emission1}) represents the 
contribution to the emission coefficient coming from collisionally excited 
atoms. 
Since collisions are assumed isotropic, this term only contributes to 
Stokes-$I$.

As an example suitable to illustrating the sensitivity of the emergent 
scattering line polarization to the studied collisional rates, we consider a 
constant-property slab of stellar atmospheric plasma located at a given 
height above the surface of a solar-like star and characterized by a 
given optical depth $\Delta \tau$. 
Neglecting limb-darkening effects, the radiation illuminating the slab from 
below is characterized by an anisotropy factor $w=\sqrt{2} \, J^2_0/ J^0_0$ 
given by 
\begin{equation}
	w = \frac{\cos\alpha \, (1 + \cos\alpha)}{2} \; ,
\end{equation}
where $\alpha$ is half the angle subtended by the stellar disk, as seen from 
the slab. 
We solve the equations of the non-LTE problem described in this section for 
the case of a $^2{\rm S}{\,}-{\,}^2{\rm P}$ transition, using the level
energies and transition probabilities of the Mg~{\sc ii} $h$ and $k$ 
lines, and we calculate the fractional linear polarization of the radiation 
emerging at $\mu=\cos\theta=0.1$, with $\theta$ the angle formed by the local 
vertical (perpendicular to the slab) and the emission direction.
The non-LTE radiative transfer problem is solved following the numerical 
methods described in \citet{JTB99}.
\begin{figure}[!t]
\centering
\includegraphics[width=\textwidth]{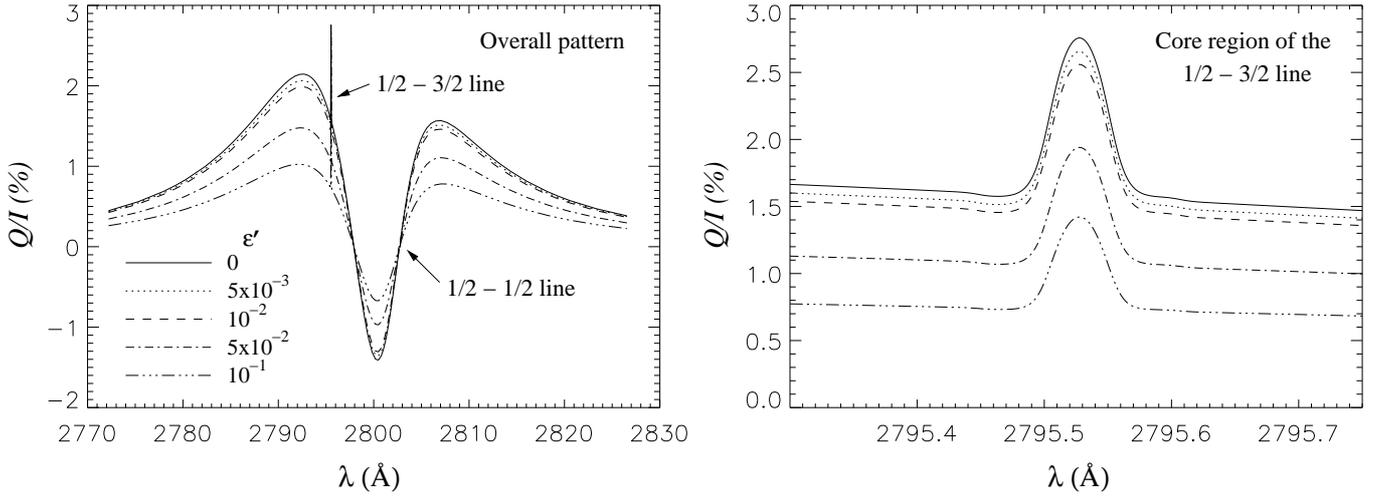}
\caption{Left panel: $Q/I$ profile of the radiation emitted across a
$^2{\rm S}{\,}-{\,}^2{\rm P}$ transition (the level energies and transition 
probabilities are those of the Mg~{\sc ii} $h$ and $k$ lines) as obtained for 
different values of the parameter $\epsilon^{\prime}$ (indicated in the plot). 
The arrows point to the wavelength positions of the two lines.
Right panel: zoom of the line-core region of the $1/2 - 3/2$ transition. 
We consider the radiation emitted at $\mu=0.1$ by a slab located 0.03 stellar 
radii above the surface, and with an optical depth (at the line-center 
frequency of the $1/2 - 1/2$ transition) $\Delta \tau = 0.5$.
We solve the full non-LTE radiative transfer problem within the slab, the 
boundary condition being the stellar radiation illuminating the slab from below 
(limb-darkening effects are neglected).
We consider a Doppler width of 26~m{\AA}, corresponding to a temperature of 
$10^4$~K and a microturbulent velocity of 1~km/s.
We include the effect of an unpolarized continuum characterized by an opacity 
$10^8$ times less than the line opacity at the line-center frequency of the 
$1/2 - 1/2$ transition. 
The reference direction for positive $Q$ is the parallel to the closest limb.}
\end{figure}

Figure~1 shows the fractional linear polarization $Q/I$ pattern calculated for 
different values of $\epsilon^{\prime}$, considering a slab located 0.03 
stellar radii above the surface (corresponding to about $2 \times 10^4$~km 
in the solar case), and characterized by an optical depth (at the line-center 
frequency of the $1/2 - 1/2$ transition) $\Delta \tau=0.5$.
We assume a Doppler width of 26~m{\AA}, corresponding to a temperature of 
$10^4$~K, and a microturbulent velocity of 1~km/s. 
The damping constant is consistently calculated as
\begin{equation}
	a=\frac{\Gamma}{\Delta \nu_D} = \frac{\gamma_u}{4 \pi \Delta \nu_D} \; ,
\end{equation}
with $\gamma_u = A(\beta_u L_u S \rightarrow \beta_{\ell} L_{\ell} S) +
\mathcal{C}_S(\beta_u L_u S \rightarrow \beta_{\ell} L_{\ell} S)$ the inverse 
lifetime of the upper term (elastic collisions, hence their broadening 
effect, are neglected).
We include the contribution of an unpolarized continuum characterized by 
an opacity $\eta_I^c = 10^{-8} \, \eta_I^{\ell}(\nu_{1/2-1/2})$, with 
$\eta_I^{\ell}(\nu_{1/2-1/2}) = k_M^A \, \varphi(\nu_{1/2-1/2})$ the line 
opacity at the line-center frequency of the $1/2-1/2$ transition.
We first note that in the slab model that we have considered (in which 
radiative transfer effects are significant), the $Q/I$ profiles show the 
typical signatures of $J$-state interference, such as the sign-reversal 
between the two lines, and the high polarization values in the far wings (see 
Stenflo 1980, LL04, and Belluzzi \& Trujillo Bueno 2011).
As can be observed, the modification of the scattering line polarization  
pattern due to inelastic and superelastic collisions ({\it quenching effect}) 
becomes appreciable only for rather high values of the collisional rates 
($\epsilon^{\prime} {\gtrsim} 10^{-2}$).

\section{Conclusions}
In this paper we have formally defined and investigated the transfer and 
relaxation rates due to isotropic inelastic and superelastic collisions that 
enter the statistical equilibrium equations for the atomic density matrix of 
a multiterm atom (i.e., a model atom accounting for quantum interference 
between magnetic sublevels pertaining either to the same $J$-level, or to 
different $J$-levels within the same term).

While the numerical values of the collisional rates for $J$-level populations 
are generally available (either from approximate theoretical expressions or 
form experimental data), the values of the collisional rates describing the 
transfer and relaxation of quantum coherence are in most cases unknown.
In this work we focused our attention on the collisional rates for $J$-state 
interference (the physical aspect that cannot be accounted for with a 
multilevel model atom).
Under the assumption that the interaction between the atom and the perturber is 
described by a dipolar operator, we derived suitable relations between such 
rates and the usual collisional rates for $J$-level populations.
In particular, we showed that the collisional relaxation rate for $J$-state 
interference coincide with the relaxation rate for $J$-level populations 
and for interference between magnetic sublevels pertaining to the same 
$J$-level. We also observed that this rate does not depend on the particular 
$J$-level under consideration, so that it is sufficient to 
introduce a single collisional relaxation rate for the whole term.
As a consistency proof of our derivations, we showed that the transfer and 
relaxation rates for $J$-level populations and for interference between pairs 
of magnetic sublevels pertaining to the same $J$-level reduce to those derived 
in Sect.~7.13 of LL04 for the multilevel atom case. 

As an illustrative application, we considered a constant-property slab of 
given optical depth, located at a given height above the surface of a 
solar-like star, and anisotropically illuminated by its photospheric radiation 
field.
The numerical solution of the full non-LTE problem for the case of a two-term 
atom with unpolarized lower term shows that the polarization of the radiation 
emerging from the slab at $\mu=0.1$ is sensitive to the presence of isotropic 
inelastic and superelastic collisions only for values of the parameter 
$\epsilon^{\prime}= \mathcal{C}_{S}(\beta_u L_u S \rightarrow \beta_{\ell} 
L_{\ell} S) / A(\beta_u L_u S \rightarrow \beta_{\ell} L_{\ell} S)$ on the 
order of $10^{-2}$ or larger.
Such values are actually needed to produce an appreciable variation in the 
$Q/I$ profiles of the emergent radiation.

\begin{acknowledgements}
Financial support by the Spanish Ministry of Science through projects
AYA2010-18029 (Solar Magnetism and Astrophysical Spectropolarimetry) and
CONSOLIDER INGENIO CSD2009-00038 (Molecular Astrophysics: The Herschel and
Alma Era) is gratefully acknowledged.
\end{acknowledgements}

\newpage

\appendix

\section{Properties of 3-$j$ and 6-$j$ symbols and of rotation matrices}
In this appendix we recall some useful properties and relations of 3-$j$ and 
6-$j$ symbols, as well as of rotation matrices that are used in the derivation of the 
expressions presented in the paper.
A proof of these properties can be found in Chapter~2 of LL04.

\begin{itemize}
	\item{Symmetry properties of 3-$j$ symbols:
		
		The 3-$j$ symbols are invariant under cyclic permutations of 
		their columns and are multiplied by $(-1)^{a+b+c}$ under 
		noncyclic ones
		\begin{equation}
			\left( \begin{array}{ccc}
				a & b & c \\
				\alpha & \beta & \gamma 
			\end{array} \right) =
			\left( \begin{array}{ccc}
				b & c & a \\
				\beta & \gamma & \alpha
			\end{array} \right) =
			(-1)^{a+b+c}
			\left( \begin{array}{ccc}
				c & b & a \\
				\gamma & \beta & \alpha
			\end{array} \right) \; , \;\; {\rm etc.}
		\label{Eq:3ja}
		\end{equation}
		
		The 3-$j$ symbols are multiplied by $(-1)^{a+b+c}$ under sign 
		inversion of the second row
		\begin{equation}
			\left( \begin{array}{ccc}
				a & b & c \\
				\alpha & \beta & \gamma 
			\end{array} \right) =
			(-1)^{a+b+c}
			\left( \begin{array}{ccc}
				a & b & c \\
				-\alpha & -\beta & -\gamma
			\end{array} \right) \; .
		\label{Eq:3jb}
		\end{equation}}

	\item{Orthogonality relation of 3-$j$ symbols
		\begin{equation}
			\sum_{\alpha \beta} (2c +1)
			\left( \begin{array}{ccc}
				a & b & c \\
				\alpha & \beta & \gamma 
			\end{array} \right)
			\left( \begin{array}{ccc}
				a & b & c^{\prime} \\
				\alpha & \beta & \gamma^{\prime}
			\end{array} \right) =
			\delta_{c c^{\prime}} \, \delta_{\gamma \gamma^{\prime}}
			\; .
		\label{Eq:3j_orto}
		\end{equation}}

	\item{Analytical expression of 3-$j$ symbols for particular values of 
		the arguments:
		\begin{equation}
			\left( \begin{array}{ccc}
				a & b & 0 \\
				\alpha & \beta & 0
			\end{array} \right) =
			(-1)^{a-\alpha} \, \delta_{ab} \, 
			\delta_{\alpha, -\beta} \, \frac{1}{\sqrt{2a +1}} \; .
		\label{Eq:3jc}
		\end{equation}}

	\item{Symmetry properties of 6-$j$ symbols:
		
		The 6-$j$ symbols are invariant under interchange of any two 
		columns and under interchange of the upper and lower arguments 
		in any two columns.}

	\item{Sum rules of 6-$j$ symbols:
\begin{align}
	\sum_c (2c +1)(2f +1)
	\left\{ \begin{array}{ccc}
		a & b & c \\
		d & e & f
	\end{array} \right\}
	\left\{ \begin{array}{ccc}
		a & b & c \\
		d & e & g
	\end{array} \right\} 
	= & \delta_{fg} 
	\label{Eq:6j_sum}
	\; , \\
	\sum_c (-1)^{a+b+c+d+e+f+g+h+i+j}(2c +1)
	\left\{ \begin{array}{ccc}
		a & b & c \\
		d & e & f
	\end{array} \right\}
	\left\{ \begin{array}{ccc}
		a & b & c \\
		g & h & i
	\end{array} \right\} 
	\left\{ \begin{array}{ccc}
		g & h & c \\
		e & d & j
	\end{array} \right\} = & 
	\left\{ \begin{array}{ccc}
		f & i & j \\
		g & d & b
	\end{array} \right\} 
	\left\{ \begin{array}{ccc}
		f & i & j \\
		h & e & a
	\end{array} \right\} \; .
	\label{Eq:6j_sum2}
\end{align}}

	\item{Analytical expression of 6-$j$ symbols for particular values 
		of the arguments
\begin{equation}
	\left\{ \begin{array}{ccc}
		a & b & 0 \\
		d & e & f
	\end{array} \right\} =
	\delta_{a b} \, \delta_{e d} \, (-1)^{a + e + f}
	\frac{1}{\sqrt{(2a + 1)(2d + 1)}} \; .
\label{Eq:6ja}
\end{equation}}

	\item{Contraction of 3-$j$ symbols:
\begin{equation}
	\sum_f (-1)^{a+b+c+d-e+f-\alpha-\delta} (2f+1)
	\left\{ \begin{array}{ccc}
		a & b & e \\
		d & c & f
	\end{array} \right\} 
	\left( \begin{array}{ccc}
		c & a & f \\
		\gamma & \alpha & \phi
	\end{array} \right)
	\left( \begin{array}{ccc}
		b & d & f \\
		\beta & \delta & -\phi
	\end{array} \right) =
	\left( \begin{array}{ccc}
		a & b & e \\
		\alpha & \beta & -\epsilon
	\end{array} \right)
	\left( \begin{array}{ccc}
		d & c & e \\
		\delta & \gamma & \epsilon
	\end{array} \right) \; .
\label{Eq:3j_contr}
\end{equation}}
	
	\item{Orthogonality relations of rotation matrices
		\begin{equation}
			\sum_P \mathcal{D}^{J}_{P N}(R)^{\ast}
			\mathcal{D}^{J}_{P M}(R) =
			\delta_{M N} \; .
		\label{Eq:rotmat_orto}
		\end{equation}}
	
	\item{Product of two rotation matrices:
\begin{equation}
	\mathcal{D}^J_{MN}(R) \,
	\mathcal{D}^{J^{\prime}}_{M^{\prime} N^{\prime}}(R)^{\ast} = 
	(-1)^{M^{\prime} - N^{\prime}} \sum_K (2K +1) 
	\left( \begin{array}{ccc}
		J & J^{\prime} & K \\
		M & -M^{\prime} & Q
	\end{array} \right)
	\left( \begin{array}{ccc}
		J & J^{\prime} & K \\
		N & -N^{\prime} & Q^{\prime}
	\end{array} \right)
	\mathcal{D}^{K}_{Q Q^{\prime}}(R)^{\ast} \; ,
\label{Eq:rotmat1}
\end{equation}
\begin{equation}
	\mathcal{D}^J_{MN}(R) \,
	\mathcal{D}^{J^{\prime}}_{M^{\prime} N^{\prime}}(R) = 
	\sum_K (2K +1) 
	\left( \begin{array}{ccc}
		J & J^{\prime} & K \\
		M & M^{\prime} & Q
	\end{array} \right)
	\left( \begin{array}{ccc}
		J & J^{\prime} & K \\
		N & N^{\prime} & Q^{\prime}
	\end{array} \right)
	\mathcal{D}^{K}_{Q Q^{\prime}}(R)^{\ast} \; .
\label{Eq:rotmat2}
\end{equation}}
\end{itemize}

\end{document}